\documentclass[12pt,british]{iopart}
\usepackage[T1]{fontenc}
\usepackage[latin9]{inputenc}
\usepackage{units}
\usepackage{graphicx}

\makeatletter
\usepackage{iopams}
\usepackage{setstack}

\makeatother

\usepackage{babel}
\begin{document}

\title{Negative probabilities and counter-factual reasoning in quantum cognition}

\author{J. Acacio de Barros$^{1,*}$, and G. Oas$^{2,\dagger}$}

\address{$^{1}$ LS Program, 1600 Holloway Street, San Francisco State University,
San Francisco, CA 94132}

\address{$^{2}$ SPCS, 220 Panama Street, Stanford University, Stanford, CA
94305-4101}

\ead{$^{*}$barros@sfsu.edu, $^{\dagger}$oas@stanford.edu}
\begin{abstract}
In this paper we discuss quantum-like decision-making experiments
using negative probabilities. We do so by showing how the two-slit
experiment, in the simplified version of the Mach-Zehnder interferometer,
can be described by this formalism. We show that negative probabilities
impose constraints to what types of counter-factual reasoning we can
make with respect to (quantum) internal representations of the decision
maker. 
\end{abstract}


\maketitle

\section{Introduction}

The use of the mathematical apparatus of quantum mechanics in the
social sciences has experienced a boom in recent years, with whole
books solely devoted to certain aspects of it (see \cite{khrennikov_ubiquitous_2010,busemeyer_quantum_2012,haven_quantum_2013}
and references therein). Such an endeavour has been dubbed ``quantum
interactions,'' and it has been the topic of many academic meetings,
such as the special section in Andrei Khrennikov's foundations of
quantum mechanics conference. The goal of the quantum interaction
community is not to search for ``true'' quantum mechanical effects,
but instead to ask whether there are social phenomena that could be
better described by, say, vectors on a Hilbert space and Hermitian
operators representing observable quantities. 

In many cases the quantum-like probabilistic structure is modelled
by introducing the use of ``quantum'' intermediate states, often
not accessible under certain experimental conditions, whose superposition
are similar to the superposition of the two quantum waves, $\psi_{A}$
and $\psi_{B}$, in analogy to the two-slit experiment. This is, for
instance, how the violation of Savage's Sure-Thing Principle (STP)
is modelled with state vectors \cite{busemeyer_quantum_2006}. Thus,
it should come as no surprise that one of the often used analogies
is interference in the two-slit experiment, since it is perhaps best
illustrates some of the puzzling quantum aspects \cite{khrennikov_ubiquitous_2010,busemeyer_quantum_2012,haven_quantum_2013,aerts_quantum_2007,khrennikov_quantum_2009,khrennikov_quantum-like_2009}. 

The importance of the two slit in physics is emphasized by Richard
Feynman in his celebrated lectures on physics \cite{feynman_feynman_2011}.
There, Feynman states that the \emph{only} mystery in quantum mechanics
is contained in this experiment, and that it ``is impossible, absolutely
impossible, to explain in any classical way'' this phenomenon. Though
grossly inaccurate%
\footnote{For example, as Feynman was well aware, David Bohm provided a way
to think about the two slit in terms of classical concepts of waves
and particles \cite{bohm_suggested_1952-1}. %
}, Feynman's claim carries a certain element of truth, as the outcomes
of the two slit experiment seem to make no sense if we think solely
in terms of particles and detectors. At the heart of such mystery
is connection between the particle's trajectory and its surroundings
(the context), yielding probabilistic outcomes that are not only context-dependent
but also non-monotonic. 

The above ``mystery'' raises the prospect of thinking of quantum-like
effects in the social sciences as originated by the interference effects
of highly contextual variables \cite{khrennikov_quantum-like_2004,de_barros_quantum_2009}.
For instance, in the case of human decision making, one can show that,
under reasonable assumptions, neural oscillators in the brain may
lead to interference-like effects that are similar to a two-slit set-up
\cite{de_barros_quantum-like_2012,suppes_phase-oscillator_2012,de_barros_response_2014}.
This interference is highly contextual, as it depends on boundary
conditions on the dynamics of neural oscillators, and can not be described
by standard probability theory. Furthermore, such contextual variables
may have a description in terms of extended probabilities that is
more general than the standard quantum formalism \cite{de_barros_joint_2012,de_barros_decision_2014,oas_exploring_2014}. 

Part of the mystery is related to the use of counter-factual reasoning
\cite{laloe_we_2001,vaidman_counterfactuals_2009}. Counter-factual
reasoning in quantum mechanics is based on non-measured properties
of a system. Suppose that we have two observations $\mathbf{A}=1$
and $\mathbf{B}=1$, with $\mathbf{A}$ preceding (perhaps even causally)
$\mathbf{B}$. In this case, a counter-factual reasoning would be
something like ``if \textbf{$\mathbf{A}=1$} had not been observed,
$\mathbf{B}=1$ would not have been observed.\textquotedblright{}
It is often pointed out that the inconsistencies associated with quantum
mechanics come from the use of counter-factual reasoning. For example,
in the famous GHZ paradox \cite{greenberger_going_1989}, inconsistencies
are derived from the assumption that values of some observables are
defined even if experimental conditions when they had not been observed,
which is equivalent to requiring the existence of a non-negative joint
probability distribution \cite{de_barros_inequalities_2000}.

The use of negative probabilities connected to the quantum formalism
dates back from the seminal works of Eugene Wigner \cite{wigner_quantum_1932},
where he showed that if one wanted a joint probability distribution
of momentum and position that would reproduce quantum statistical
mechanics predictions, such joint would have to take negative values.
Later on, Paul Dirac \cite{dirac_bakerian_1942} proposed the use
of negative probabilities to understand the quantization of electromagnetic
fields. However, as Richard Feynman remarked, even though negative
probabilities seem to be a part of quantum mechanics, no practical
use was ever found for them \cite{feynman_negative_1987}, although
recently some authors proposed possible applications \cite{oas_exploring_2014,abramsky_operational_2014,al-safi_simulating_2013}. 

Perhaps one of the difficulties of using negative probabilities is
its unclear physical meaning. Wigner considered them nonsensical,
and both Dirac and Feynman thought they were mere computational devices
devoid of meaning. However, recently many different interpretations
have appeared. For example, in \cite{de_barros_decision_2014} a possible
subjective interpretation in terms of non-monotonic upper probability
distributions was proposed, and in \cite{abramsky_operational_2014}
a map based on positive and negative measures gave a possible frequentist
interpretation to negative probabilities. Another approach was that
of Khrennikov, who showed that negative probabilities are obtained
in the context of $p$-adic statistics, and are the consequence of
a violation of principle of statistical stabilization \cite{khrennikov_p-adic_1993,khrennikov_p-adic_1993-1,khrennikov_p-adic_1994,khrennikov_discrete_1994,khrennikov_statistical_1993}.
Be that as it may, negative probabilities in quantum mechanics seem
to be related to the appearance of correlations that are too strong
to be explained by a context-independent hidden-variable theory \cite{de_barros_unifying_2014},
such as the case for the well-known Bell-EPR systems. 

Our goal in this paper is to examine the use of extended probabilities
in the context of the two-slit experiment, as suitable for discussions
in quantum cognition. In particular, we examine this experiment in
its simplified version: the Mach-Zehnder interferometer. We then discuss
what constraints are imposed to a rational description of the system
when further (counter-factual) information is assumed about the system.
To do so, we organize this paper in the following way. In Section
\ref{sec:Interference-in-quantum} we discuss a typical application
of the quantum formalism in decision making, showing its equivalence
to the quantum Mach-Zehnder interferometer. In Section \ref{sec:Counterfactuals-and-signed}
we analyse in terms of negative probabilities the Mach-Zehnder set-up,
and show that certain types of counter-factual reasoning are inconsistent
with them.

\section{Interference in Quantum Cognition\label{sec:Interference-in-quantum}}

Let us start with the famous violation of STP, observed empirically
by Tversky and Shafir \cite{shafir_thinking_1992,tversky_disjunction_1992}.
In one of their experiments, participants (students in a college-level
class) were asked whether they would buy a certain ticket under three
different contexts. In the first context, they were asked if they
would buy if the passed the class. In the second context, they were
asked if they would buy if they failed the class. Finally, in the
third context they were asked whether they would buy the ticket if
they didn't know whether they passed the class or not. Participants
consistently preferred to buy the tickets when they knew whether they
passed or failed the class than if they didn't know, a violation of
STP. 

To understand this violation of STP, it helps to write the situation
in a more formal way. Let $\mathbf{P}$ and $\mathbf{B}$ be two $\pm$1-valued
random variables. $\mathbf{B}$ represents the intent to buy a ticket,
and $\mathbf{P}$ represents passing the class, with both taking value
$1$ if the answer was ``yes'' and $-1$ if ``no''. The probability
of purchasing a ticket in the first context corresponds to the value
$P\left(b|p\right)$, the second context to $P\left(b|\overline{p}\right)$.
Since in the third context the participant does not know, but $1$
and $-1$ are the only possible outcomes, the third context must be
$P\left(b|p\cup\overline{p}\right)=P\left(b\right)$. Here we use
the standard notation that $b$ represents $\mathbf{B}=1$, $\overline{b}$
represents $\mathbf{B}=-1$, and so on. Since the probability of $\mathbf{B}=1$
is given by 
\begin{equation}
P\left(b|p\cup\overline{p}\right)=P\left(b\right)=P\left(b|p\right)P\left(p\right)+P\left(b|\overline{p}\right)P\left(\overline{p}\right),\label{eq:}
\end{equation}
with $P\left(p\right)+P\left(\overline{p}\right)=1$, it follows that
$P\left(b|p\cup\overline{p}\right)$ is a convex combination of $P\left(b|p\right)$
and $P\left(b|\overline{p}\right)$. Thus, $P\left(b|p\cup\overline{p}\right)\geq P\left(b|p\right)$.
This, of course, contradicts the results of Tversky and Shafir, and
shows that their participants violate STP, a consequence of the standard
theory of probabilities. 

It is worthwhile mentioning here that the main characteristic of the
data, as the one discussed above, is to show that, under certain circumstances,
human decision making is non-monotonic. Recall that standard probability
theory is monotonic, in the sense that for two sets $A$ and $B$
with $B\subseteq A$, $P\left(A\right)\geq P\left(B\right)$. Monotonicity
is a consequence of the additivity rule, which gives $P\left(A\right)=P\left(B\cup\overline{B}\right)=P\left(B\right)+P\left(\overline{B}\right)$,
with $\overline{B}=A\setminus B$ being the relative complement of
$B$ in $A$, plus the non-negativity axiom. This means that certain
extended probability theories that are monotonic, such as Dempster-Shafer
\cite{dempster_upper_1967,shafer_mathematical_1976}, may not be adequate
to describe actual human decision making. 

\begin{figure}
\begin{centering}
\includegraphics{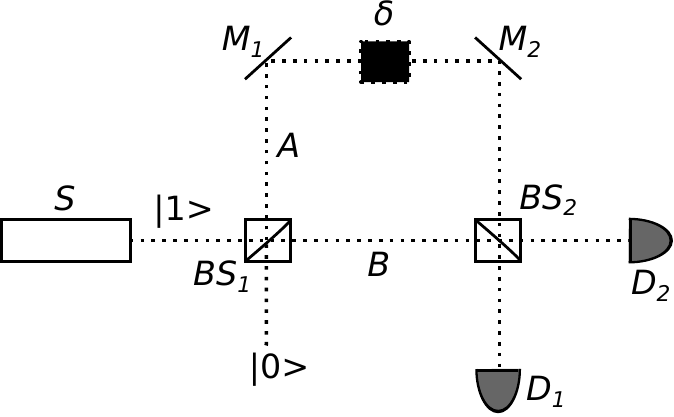}
\par\end{centering}

\protect\caption{\label{fig:Mach-Zehnder-interferometer}Mach-Zehnder interferometer.
A single particle emitted from source $S$ reaches beam-splitter $BS_{1}$,
and has a $50/50$ probability of going to either arm, $A$ or $B$,
of the interferometer. A second beam-splitter $BS_{2}$ then joints
both paths $A$ and $B$ and sends them to detectors $D_{1}$ and
$D_{2}$. The intensities seem on $D_{1}$ and $D_{2}$ depend on
the relative phases of each beam, determined by the phase shifter
$\delta$. }

\end{figure}

Non-monotonicity is directly connected to the quantum formalism via
quantum interference. To understand this, let us examine the Mach-Zehnder
interferometer, depicted in Figure \ref{fig:Mach-Zehnder-interferometer},
as a simplified two-slit experiment. 
 The input into the system is the state $|1\rangle_{S}|0\rangle_{V}$,
corresponding to one photon in channel $S$ and no photon in the vacuum
channel $V$. After the first beam splitter the state becomes 
\[
\frac{1}{\sqrt{2}}\left[|1\rangle_{A}|0\rangle_{B}+|0\rangle_{A}|1\rangle_{B}\right],
\]
where the subscripts $A$ and $B$ are the modes in the arms of the
interferometer, and the state corresponds to a superposition of a
photon in $A$ or $B$. Ignoring the reflection on the mirrors, since
what is relevant is the phase introduced by $\delta$, the state impinging
upon the second beam splitter is
\[
\frac{1}{\sqrt{2}}\left[e^{i\delta}|1\rangle_{A}|0\rangle_{B}+|0\rangle_{A}|1\rangle_{B}\right].
\]
 Finally, the state impinging upon the detectors is given by 
\[
\frac{1}{2}\left(1-e^{i\delta}\right)|1\rangle_{1}|0\rangle_{2}+\frac{1}{2}\left(1+e^{i\delta}\right)|0\rangle_{1}|1\rangle_{2},
\]
where subscripts $1$ and $2$ refer to detectors $D_{1}$ and $D_{2}$
respectively. As we can see, the probability of detections are $P\left(d_{1}\right)=\frac{1}{2}\left(1-\cos\delta\right)$
and $P\left(d_{2}\right)=\frac{1}{2}\left(1+\cos\delta\right)$, and
if we select $\delta=\pi$, the probability of detection on $D_{1}$
is one whereas on $D_{2}$ it is zero. However, if we measure the
particle in one of the interferometer arms, say after the phase $\delta$
is introduced, we end up with a mixture of two possible states, $|1\rangle_{a}|0\rangle_{b}$
or $|0\rangle_{a}|1\rangle_{b}$. After the last beam splitter, we
get the state 
\[
\frac{1}{\sqrt{2}}\left[|1\rangle_{1}|0\rangle_{2}+|0\rangle_{1}|1\rangle_{2}\right],
\]
and we see that the probabilities for detection are now $P\left(d_{1}\right)=P\left(d_{2}\right)=\frac{1}{2}$
. 

From the above computations we can now see the non-monotonicity of
this set-up. Let us imagine two possibilities. Possibility one is
such that the photon can only travel through path $A$, perhaps by
simply placing a barrier on path $B$. Possibility two is when no
barriers are placed, and the photon can freely travel through either
arm of the interferometer. According to monotonic reasoning, the more
possibilities we have going from $S$ to $D_{2}$, the higher the
probability of $D_{2}$. However, if we choose $\delta=\pi$, $P\left(d_{2}\right)=0$
for both paths being available, and $P\left(d_{2}\right)=\frac{1}{2}$
for only one path available. In other words, $P\left(d_{2}|a\cup b\right)\leq P\left(d_{2}|a\right)$,
similar to the violation of Savage's STP mentioned above. 

To model the violation of Savage's STP with a quantum formalism, all
we need to do is to think of the path information $|1\rangle_{a}$
or $|1\rangle_{b}$ as the ``pass'' or ``fail'' class intermediate
states, say $|\mbox{pass}\rangle$ and $|\mbox{fail}\rangle$. The
final detectors $D_{1}$ and $D_{2}$ corresponding to ``buy'' or
``not buy'' the ticket. So, in one experimental condition, when
the experimenter asks the answer if the participant passed the class,
this would correspond to measuring a which path the particle travelled,
thus collapsing the wave function in the $|\mbox{pass}\rangle$ state
and destroying the interference. In other words, in the version of
the experiment where participants are asked their answers if the passed
or failed, their state evolves as 
\[
|\mbox{pass}\rangle\mbox{ or }|\mbox{fail}\rangle\longrightarrow\frac{1}{\sqrt{2}}\left[|\mbox{purchase}\rangle+|\mbox{not purchase}\rangle\right],
\]
whereas in the other version the evolution is 
\[
\frac{1}{\sqrt{2}}\left[|\mbox{pass}\rangle+|\mbox{fail}\rangle\right]\longrightarrow\frac{1}{2}\left[\left(1-e^{i\delta}\right)|\mbox{purchase}\rangle+\left(1+e^{i\delta}\right)|\mbox{not purchase}\rangle\right],
\]
where $\delta$ is a phase parameter responsible for interference.

\section{Counter-factuals and Signed Probabilities\label{sec:Counterfactuals-and-signed}}

The violation of Kolmogorov's axioms can be shown to correspond, in
the case of quantum systems, to the existence of a signed probability
distribution \cite{al-safi_simulating_2013,oas_exploring_2014,abramsky_operational_2014},
where the non-negativity axiom is relaxed. As mentioned in the introduction,
negative probabilities are the consequence of correlations between
different random variables that cannot be explained by a context-independent
hidden variable \cite{de_barros_unifying_2014}. Though the correlations
themselves are observable in physical systems, negative probability
states are not observable in principle \cite{oas_exploring_2014},
whereas this may not necessarily be true for social systems where
negative probabilities appear \cite{de_barros_decision_2014,khrennikov_interpretations_2009}.
In this section we show how negative probabilities can be used in
the Mach-Zehnder case shown above, and what are the possible insights
that we may gain from it. 

Let us imagine that we perform a destructive measurement. The constraints
given by the marginals are 
\begin{equation}
P(a\overline{b})=P\left(\overline{a}b\right)=\frac{1}{2},\label{eq:aorb}
\end{equation}
 and 
\begin{equation}
P(ab)=P(\overline{a}\overline{b})=0,\label{eq:noabornotab}
\end{equation}
which corresponds to having only one photon at a time. Here we use
the notation $a$ as corresponding a detection in the interferometer
arm $A$, and $\overline{a}$ for no detection, and similarly for
$b$, $d_{1}$, and $d_{2}$. A different experiment, when no detection
is made for $A$ or $B$, thus yielding interference at the detectors,
requires 
\begin{equation}
P(d_{1}\overline{d}_{2})=1-\alpha,\label{eq:d1d2}
\end{equation}
\begin{equation}
P(\overline{d}_{1}d_{2})=\alpha,\label{eq:d1d2-2}
\end{equation}
\begin{equation}
P(d_{1}d_{2})=P(\overline{d}_{1}\overline{d}_{2})=0,\label{eq:d1notd2andothers}
\end{equation}
where $\alpha=0$ or $\alpha=1$ give maximum interferometric visibility.
A non-negative probability distributions exists for (\ref{eq:aorb})--(\ref{eq:d1notd2andothers})
if we assume that (\ref{eq:aorb}) and (\ref{eq:noabornotab}) still
holds when we have interference. For example, a possible (non-unique)
solution consistent with the marginals is given by 
\[
P\left(a\overline{b}d_{1}\overline{d}_{2}\right)=P\left(\overline{a}bd_{1}\overline{d}_{2}\right)=\frac{1}{2}\left(1-\alpha\right),
\]
\[
P\left(a\overline{b}\overline{d}_{1}d_{2}\right)=P\left(\overline{a}b\overline{d}_{1}d_{2}\right)=\frac{1}{2}\alpha,
\]
and the other atomic elements of the algebra (e.g., $abd_{1}d_{2}$,
$abd_{1}\overline{d}_{2}$, etc.) having probability zero. 

But we also know, from another experimental set-up where we place
a detector in \emph{only one} of the interferometer arms, that when
the particle is \emph{not} in $A$ and no detector in $B$, there
is a probability of detection in $D_{1}$ and $D_{2}$ given by 
\begin{equation}
P(\overline{a}d_{1}\overline{d}_{2})=P(\overline{a}\overline{d}_{1}d_{2})=P(\overline{b}d_{1}\overline{d}_{2})=P(\overline{b}\overline{d}_{1}d_{2})=\frac{1}{4},\label{eq:p2detect}
\end{equation}
which implies the destruction of interference by which-path information.
Contrary to the marginals (\ref{eq:aorb})--(\ref{eq:d1notd2andothers}),
there is no non-negative joint probability distribution consistent
with (\ref{eq:d1d2})--(\ref{eq:p2detect}). However, if we relax
the requirement that probabilities must be non-negative, it is possible
to show that there is an infinite number of (negative) joint probability
distributions that are consistent with (\ref{eq:d1d2})--(\ref{eq:p2detect}). 

Once the non-negativity requirement is relaxed, solutions consistent
with marginals can start to have very large (negative and positive)
values. To further constraint the negative probability solutions,
we add the principle that the probability mass $M$, defined as $M=\sum\left|p_{i}\right|$
($p_{i}$ being the probability for each atom), should be minimized.
Intuitively minimizing the $L1$ norm of the probability $P$ corresponds
to requiring that $P$ be as close as possible to a non-negative probability
distribution \cite{de_barros_decision_2014}, and prevents solutions
with large negative or positive values. 

The general solution consistent with (\ref{eq:d1d2})--(\ref{eq:p2detect})
and minimizing the probability mass $M$ is, for the non-zero probability
terms,
\begin{equation}
\begin{array}[t]{cc}
P\left(abd_{1}\overline{d}_{2}\right)=-\frac{3}{4}+x+\alpha, & P\left(ab\overline{d}_{1}d_{2}\right)=\frac{3}{4}-x-\alpha,\\
P\left(a\overline{b}d_{1}\overline{d}_{2}\right)=\frac{1}{2}-x, & P\left(a\overline{b}\overline{d}_{1}d_{2}\right)=x,\\
P\left(\overline{a}bd_{1}\overline{d}_{2}\right)=\frac{1}{2}-x, & P\left(\overline{a}b\overline{d}_{1}d_{2}\right)=x,\\
P\left(\overline{a}\overline{b}d_{1}\overline{d}_{2}\right)=-\frac{1}{4}+x, & P\left(\overline{a}\overline{b}\overline{d}_{1}d_{2}\right)=\frac{1}{4}-x.
\end{array}\label{eq:general-solution-minimizing}
\end{equation}
$M$ is minimized in (\ref{eq:general-solution-minimizing}) when
$\nicefrac{1}{4}\leq x\leq\nicefrac{1}{2}$ for $0\leq\alpha\leq\nicefrac{1}{4}$,
$\nicefrac{1}{4}\leq x\leq\nicefrac{3}{4}-\alpha$ for $\nicefrac{1}{4}\leq\alpha\leq\nicefrac{1}{2}$,
$\nicefrac{3}{4}-\alpha\leq x\leq\nicefrac{1}{4}$ for $\nicefrac{1}{2}\leq\alpha\leq\nicefrac{3}{4}$,
and $\nicefrac{1}{2}\leq x\leq\nicefrac{1}{4}$ for $\nicefrac{3}{4}\leq\alpha\leq1$.
Notice that if $\alpha=\nicefrac{1}{2}$, for no interference, the
joint becomes
\begin{eqnarray*}
P\left(abd_{1}\overline{d}_{2}\right)=P\left(ab\overline{d}_{1}d_{2}\right) & = & 0,\\
P\left(a\overline{b}d_{1}\overline{d}_{2}\right)=P\left(a\overline{b}\overline{d}_{1}d_{2}\right)=P\left(\overline{a}bd_{1}\overline{d}_{2}\right) & =\\
P\left(\overline{a}\overline{b}d_{1}\overline{d}_{2}\right)=P\left(\overline{a}b\overline{d}_{1}d_{2}\right)=P\left(\overline{a}\overline{b}\overline{d}_{1}d_{2}\right) & = & \frac{1}{4},
\end{eqnarray*}
which is non-negative, as expected. However, for $\alpha\neq\nicefrac{1}{2}$,
which is the case for interference, no non-negative solutions exist.
For example, in the case of maximum visibility, when $\alpha=1$,
a possible negative joint probability minimizing the probability mass
$M$ is given by (choosing $x=\nicefrac{1}{4}$) 
\[
\begin{array}[t]{cc}
P\left(abd_{1}\overline{d}_{2}\right)=\frac{1}{2}, & P\left(ab\overline{d}_{1}d_{2}\right)=-\frac{1}{2},\\
P\left(a\overline{b}d_{1}\overline{d}_{2}\right)=\frac{1}{4}, & P\left(a\overline{b}\overline{d}_{1}d_{2}\right)=\frac{1}{4},\\
P\left(\overline{a}bd_{1}\overline{d}_{2}\right)=\frac{1}{4}, & P\left(\overline{a}b\overline{d}_{1}d_{2}\right)=\frac{1}{4},\\
P\left(\overline{a}\overline{b}d_{1}\overline{d}_{2}\right)=0, & P\left(\overline{a}\overline{b}\overline{d}_{1}d_{2}\right)=0.
\end{array}
\]

On the other hand, if we use counter-factual reasoning for the above
probabilities, we have that if there is no detection in $A$, then
we must have $\mathbf{B}=1$, whereas no detection in $B$ means $\mathbf{A}=1$.
Thus, the probabilities are 
\begin{equation}
P(\overline{a}bd_{1}\overline{d}_{2})=P(\overline{a}b\overline{d}_{1}d_{2})=P(a\overline{b}d_{1}\overline{d}_{2})=P(a\overline{b}\overline{d}_{1}d_{2})=\frac{1}{4},\label{eq:p1detect-1}
\end{equation}
and all other probabilities are zero. It is easy to show that no negative
probabilities exist consistent with (\ref{eq:d1d2})--(\ref{eq:d1notd2andothers})
and (\ref{eq:p1detect-1}).

\section{Final Remarks}

We saw that in the two slit experiment, in the simpler context of
the Mach-Zehnder interferometer, proper probability distributions
exist if we do not make any \emph{assumptions} about which-path information.
However, once we include which-path information in the form of counter-factual
reasoning, no proper joint exists, and negative probabilities are
needed. Furthermore, if we assume more about the trajectory of the
particle than what is contained in (\ref{eq:p2detect}), i.e. if we
include more counter-factual assumptions as in (\ref{eq:p1detect-1}),
not even a negative probability distribution exists. 

The non-existence of a negative probability, as showed in \cite{oas_exploring_2014},
is related to the violation of the no-signalling condition, which
in psychology is equivalent to the violation of marginal selectivity
\cite{dzhafarov_selectivity_2012,dzhafarov_probabilistic_2014}. This
should not come as a surprise, because a Mach-Zehnder can be used
to signal. To see this, imagine that Alice is at the arms of the interferometer,
and Bob is looking at the detectors $D_{1}$ and $D_{2}$. Alice's
decision to measure or not where the particle is would affect the
outcomes of Bob in a way that Bob could tell whether Alice is measuring
or not. Of course, this type of signalling does not present any conceptual
problems, as it is purely classical, since it would take some time
for the particle to arrive from Alice and Bob, and their measurements
would not be space-like separated. 

Our results suggest that in quantum-like decision-making, where most
effects come from cases where the superposition of two states result
in quantum-like interference, we should expect violation of marginal
selectivity (or the no-signalling condition in physics). Furthermore,
if we want to use the empirical results to think about inaccessible
internal reasoning states (e.g., if the student answered ``buy a
ticket'', was she thinking ``passed class''?), since those states
are not directly measured, we need to use counter-factual reasoning.
However, as we showed above, not every counter-factual reasoning allows
us to obtain information about the system, as a (negative) joint probability
cannot always be found (such as in equations (\ref{eq:p1detect-1})).
Thus, to think about those internal reasoning states, one needs to
restrict the counter-factual assumptions one makes or perhaps use
a more general approach, such as the one proposed by Dzhafarov and
Kujala \cite{dzhafarov_selectivity_2012}.

\paragraph{Acknowledgement. }

The authors wish to thank Professors Patrick Suppes, Ehtibar Dzhafarov,
Janne Kujala, Stephan Hartmann, Andrei Khrennikov, Samson Abramsky,
Claudio Carvalhaes, and Jerome Busemeyer for useful discussions about
negative probabilities and quantum mechanics and psychology. We also
thank the anonymous referees for suggestions and comments.

\section*{References}

\bibliographystyle{unsrt}
\bibliography{Quantum}

\end{document}